\documentclass[%
reprint,
aps,
prb,
]{revtex4-2}
\usepackage{graphicx}
\usepackage{hyperref}

\usepackage{color}
\usepackage{xcolor}
\usepackage{multirow}
\usepackage{dcolumn}
\usepackage{bm}
\usepackage{amsmath}
\usepackage{natbib}
\usepackage{docmute}
\usepackage{amssymb}
\usepackage{ulem}
\usepackage{here}

\graphicspath{{./graph/}}

\usepackage{bm}

\begin{document}

\title{
  Generation of polarized spin-triplet Cooper pairings by magnetic barriers in superconducting junctions
}

\author{Shun Tamura$^{1}$, Yukio Tanaka$^{1}$, and Takehito Yokoyama$^{2}$}
\affiliation{%
$^1$Department of Applied Physics, Nagoya University, Nagoya 464-8603, Japan\\
$^2$Department of Physics, Tokyo Institute of Technology, Tokyo 152-8551, Japan
}

\date{\today}
\begin{abstract}
  We investigate the proximity effect in an $s$-wave superconductor/ferromagnetic metal with a Rashba spin-orbit coupling/diffusive normal metal junction and an $s$-wave superconductor/noncollinear magnetic metal/diffusive normal metal junction.
We show the generation of polarized spin-triplet pairings in the diffusive normal metal due to coherent spin rotation in the intermediate magnetic regions.
The emergence of the spin-triplet odd-frequency Cooper pairings can generate a zero energy peak in the quasiparticle density of states in the diffusive normal metal.   
\end{abstract}
\pacs{pacs}

\maketitle
\section{Introduction}
The emergence of spin-triplet pairings in superconductor (SC)/ferromagnet
junctions has received much attention~\cite{RevModPhys.77.935,RevModPhys.77.1321,doi:10.1063/1.3541944,Linder2015}.
In ferromagnet/SC junctions, 
polarized spin-triplet pairings
can be generated due to coherent spin rotation by inhomogeneous
magnetization~\cite{PhysRevLett.86.4096,PhysRevB.64.134506}.  The generation of the polarized spin-triplet pairing has
been confirmed by observing Josephson current through strong
ferromagnets~\cite{Keizer2006,PhysRevLett.104.137002,doi:10.1126/science.1189246}.
Polarized spin-triplet pairings are an important ingredient for superconducting 
spintronics. For example, in current-biased ferromagnetic Josephson junctions, one can realize spin-polarized supercurrent due to the generation of polarized spin-triplet pairings~\cite{Eschrig_2015}. Since polarized spin-triplet pairings have a spin polarization, they can also be used to exert spin transfer torques and induce magnetization dynamics~\cite{PhysRevB.65.054407,PhysRevB.83.012501,PhysRevB.90.054504,PhysRevB.98.014521,PhysRevB.96.121203}. 

Even in uniform ferromagnets, polarized spin-triplet
pairings can be generated by spin-orbit coupling due to 
coherent spin rotation~\cite{PhysRevLett.110.117003,PhysRevB.89.134517,PhysRevB.95.184508,PhysRevB.102.100507}. 
The interplay between spin-orbit coupling and superconductivity leads to various phenomena such as a zero energy peak in the density of states~\cite{PhysRevB.92.024510,Pablo2015}, $\phi_0$
junctions~\cite{PhysRevLett.101.107005,PhysRevB.92.125443,TYN2009}, magnetoelectric effects~\cite{PhysRevB.96.060502,10.1088/1361-648X/ac7994}, and
enhanced spin pumping~\cite{Jeon2018}.

Spin-triplet pairings are also generated with the use of ferromagnetic insulators. 
They are generated from spin-flip scattering by inserting a ferromagnetic insulator at the interface between a normal metal and an SC~\cite{PhysRevLett.102.107008,PhysRevB.81.214504,PhysRevB.104.054519} or by placing an SC on the ferromagnet insulator EuS~\cite{PhysRevMaterials.1.054402,PhysRevB.100.184501,PhysRevResearch.3.023131}
Although this spin-flip scattering by homogeneous ferromagnets generates spin-triplet pairings, a Cooper pair amplitude does not have polarized spin-triplet components: the Cooper pair amplitude is not polarized since the $d$ vector of the spin-triplet Cooper pair amplitude is parallel to the magnetization of the ferromagnet.
Then, an inhomogeneous spin structure, including a combination of a homogeneous spin structure and a spin dependent coupling, e.g., spin-orbit coupling, is necessary to induce polarized spin-triplet pairing.
Recent experiments on superconducting tunnel junctions with magnetic insulators GdN and EuS have indicated the emergence of odd-frequency spin-triplet pairings~\cite{Pal2017,Diesch2018}, which manifests as a zero-energy peak in the local density of states~ \cite{PhysRevLett.98.107002,PhysRevLett.98.077003,PhysRevB.75.134510,odd1,tanaka12}. 
Also, it has been predicted that the coupling between a magnon in a ferromagnetic insulator and Cooper pairs can lead to a magnon spin current~\cite{PhysRevLett.127.207001} and the formation of magnon-cooparons~\cite{arxiv2203.05336}.

In this paper, we extend the previous works that studied the generation of spin-triplet Cooper pairings in superconducting junctions with a uniform magnetic interface~\cite{PhysRevLett.102.107008,PhysRevB.81.214504,PhysRevB.104.054519} to junctions with more complicated magnetic (spin) structures~\cite{PhysRevB.105.174507}.  
Although most of the previous studies about generating the polarized spin-triplet Cooper pair amplitude were based on bulk ferromagnet junctions, we show that it is induced only by the interface complex spin structure. 
Here, we emphasize that the interface considered in Refs.~\cite{PhysRevLett.102.107008,PhysRevB.81.214504,PhysRevB.104.054519} is a homogeneous ferromagnet or magnetic impurity, and spin-triplet Cooper pairs are not polarized in these junctions.
Also, we utilize a tight-binding model in which we can choose an arbitrary value of the exchange field beyond the quasi classical theory of superconductivity, where the magnitude of the exchange field is restricted to the order of the SC gap function.
We consider two kinds of junctions: an $s$-wave SC/ferromagnet with a Rashba spin-orbit coupling (RSOC)/diffusive normal metal (DN) junction and an $s$-wave SC/noncollinear magnet/DN junction.
We clarify the generation of polarized spin triplet pairings in the DN due to coherent spin rotation in the magnetic regions. The emergence of these odd-frequency spin-triplet pairings manifests as a zero-energy peak in the local density of states.

The organization of this paper is as follows. In Sec.~\ref{sec:MM}, 
we explain our model and the method of theoretical calculations. 
We show the numerically calculated 
results in Sec.~\ref{sec:results}. We summarize our results in Sec.~\ref{sec:summary}. 

\section{Model and method\label{sec:MM}}
We consider two systems: the spin-singlet $s$-wave SC/ferromagnetic metal with Rashba spin-orbit coupling (FR)/DN junction (Sec.~\ref{sec:SC_FR_DN})
and the spin-singlet $s$-wave SC/noncollinear ferromagnetic metal (NCF)/DN junction (Sec.~\ref{sec:SC_NC_DN}).

\subsection{SC/FR/DN junction\label{sec:SC_FR_DN}}
\begin{figure}[htbp]
   \includegraphics[width=0.45\textwidth]{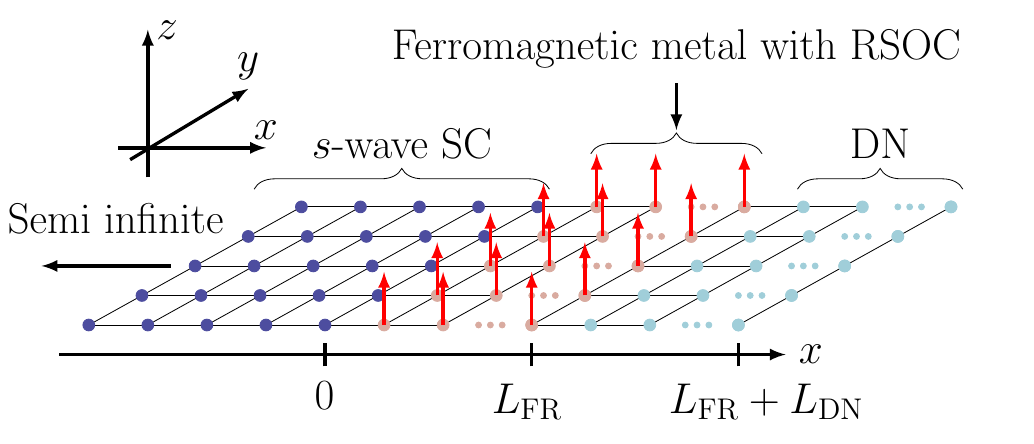}
   \caption{%
      Schematic picture of the $s$-wave SC/FR/DN junction.
      RSOC denotes Rashba spin-orbit coupling.
      We impose the periodic boundary condition in the $y$ direction. 
      In the negative $x$ direction, the SC is semi-infinite, and for the positive $x$ direction,
      we impose the open boundary condition at $j_x=L_\mathrm{FR}+L_\mathrm{DN}$.
   }
   \label{fig:Schematic_S_FR_DN_junc}
\end{figure}

The Hamiltonian for the two-dimensional SC/FR/DN junction on a two-dimensional square lattice (Fig.~\ref{fig:Schematic_S_FR_DN_junc}) is 
\begin{align}
   H_l
   =&
   H_t+
   H_\mathrm{SC}+H_\mathrm{FR}+H_{\mathrm{DN},l}^{L_\mathrm{NCF}},
   \\
   H_t
   =&
   -t\sum_{\langle \mathbf{i},\mathbf{j}\rangle,i_x,j_x\leq L_\mathrm{FR}+L_\mathrm{DN},\sigma}
   (c_{\mathbf{i},\sigma}^{\dagger} c_{\mathbf{j},\sigma}+\mathrm{H.c.}),
   \label{eq:kin_FR}
   \\
   H_\mathrm{SC}
   =&
   -\mu_\mathrm{SC}\sum_{j_x\leq0,j_y,\sigma}n_{\mathbf{j},\sigma}
   +
   \Delta\sum_{j_x\leq 0,j_y}
   (c_{\mathbf{j},\uparrow}^{\dagger} c_{\mathbf{j},\downarrow}^\dagger+\mathrm{H.c.}),
   \\
   H_\mathrm{FR}
   =&
   \sum_{1\leq j_x\leq L_\mathrm{FR},j_y,\alpha,\beta}
   \left[
      h{(\hat{\sigma}_z)}_{\alpha,\beta}-\mu_\mathrm{FR}{(\hat{\sigma}_0)}_{\alpha,\beta}
   \right]c_{\mathbf{j},\alpha}^\dagger c_{\mathbf{j},\beta}
   \nonumber\\
      &+
   i\lambda\sum_{1\leq j_x\leq L_\mathrm{FR},j_y,\alpha,\beta}
   \left[
      c_{\mathbf{j},\alpha}^\dagger{(\hat{\sigma}_y)}_{\alpha,\beta}c_{\mathbf{j}+\mathbf{e}_x,\beta}-\mathrm{H.c.}
   \right]
   \nonumber\\
      &+
   \lambda\sum_{1\leq j_x\leq L_\mathrm{FR},j_y,\alpha,\beta}
   \left[
      c_{\mathbf{j},\alpha}^\dagger{(\hat{\sigma}_x)}_{\alpha,\beta}c_{\mathbf{j}+\mathbf{e}_y,\beta}+\mathrm{H.c.}
   \right],
   \\
   H_{\mathrm{DN},l}^{L_\mathrm{FR}}
    =&
    \sum_{L_\mathrm{FR}< j_x\leq L_\mathrm{FR}+L_\mathrm{DN},j_y,\sigma}
    (V_{\mathbf{j},l}-\mu_\mathrm{DN})n_{\mathbf{j},\sigma},
    \label{eq:H_DN}
\end{align}
with $n_{\mathbf{j},\sigma}=c_{\mathbf{j},\sigma}^\dagger c_{\mathbf{j},\sigma}$.
Here, $c_{\mathbf{j},\sigma}$ ($c_{\mathbf{j},\sigma}^\dagger$) is an annihilation (creation) operator on the $\mathbf{j}$-th site with spin $\sigma$, 
$t$ is a hopping integral, $\mu_\mathrm{SC}$ is a chemical potential in the $s$-wave SC region, $\Delta$ is an $s$-wave pair potential,
$h$ is an exchange field, $\mu_\mathrm{FR}$ is a chemical potential in the FR region, $\lambda$ is a Rashba spin-orbit coupling, $V_{\mathbf{j},l}$ is an impurity potential in the DN region, $\mu_\mathrm{DN}$ is a chemical potential in the DN region, and $\hat{\sigma}_{0,x,y,z}$ is a Pauli matrix in the spin space.
We use a lattice constant as a unit of length, $\langle\mathbf{i},\mathbf{j}\rangle$ in Eq.~\eqref{eq:kin_FR} denotes the sum of nearest-neighbor pairs, and  $\mathbf{e}_{x(y)}$ denotes the unit vector in the $x$ $(y)$ direction.
As a random potential $V_{\mathbf{j},l}$, we use a uniformly distributed random number ranging from $-t$ to $t$ for each $\mathbf{j}$ and $l$~\cite{TamuraProximity,Asano2013}.
The index $l$ denotes the $l$-th impurity sample. To calculate physical quantities, we averaged over the impurity samples from $l=1$ to $l=N_\mathrm{sample}$.
We impose the periodic boundary condition in the $y$ direction with $L_y$ sites and the open-boundary condition at $j_x=L_\mathrm{FR}+L_\mathrm{DN}$ in the $x$ direction. 

Here, we consider two-dimensional systems since the computational cost of the impurity sample average is too high for more than two-dimensional systems.
We expect that the qualitative results do not change: we can obtain a polarized spin-triplet Cooper pair for three-dimensional systems.
It is noted that for three-dimensional systems, the spin-orbit coupling is not necessarily of the Rashba type, but it can be isotropic spin-orbit coupling.

\subsection{SC/NCF/DN junction\label{sec:SC_NC_DN}}
\begin{figure}[H]
   \includegraphics[width=0.45\textwidth]{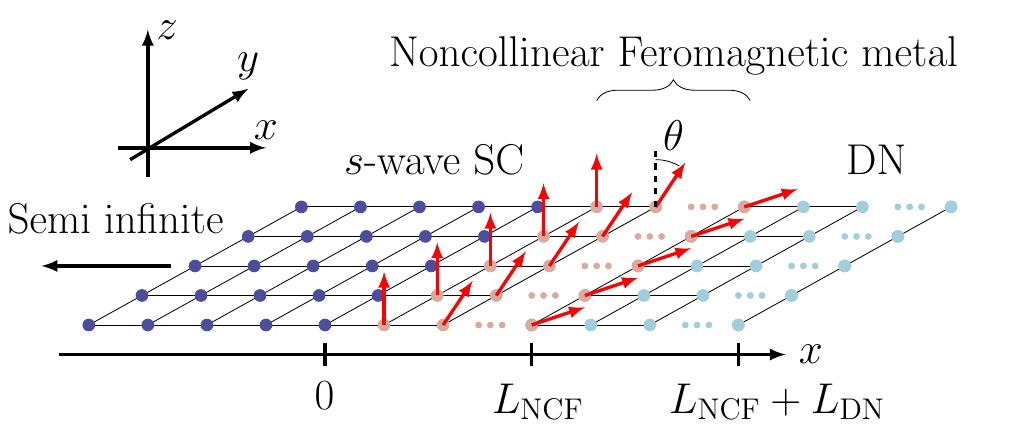}
   \caption{Schematic picture of the $s$-wave SC/NCF/DN junction.  }
   \label{fig:Schematic_S_NCx_DN_junc}
\end{figure}
The Hamiltonian for the two-dimensional SC/NCF/DN junction [Fig.~\ref{fig:Schematic_S_NCx_DN_junc}] is 
\begin{align}
   H_l =& H_t + H_\mathrm{SC} + H_{\mathrm{NCF}} + H_{\mathrm{DN},l}^{L_\mathrm{NCF}},
   \label{eq:H_NCF}
\end{align}
with 
\begin{align}
   H_{\mathrm{NCF}} =&
   \sum_{1\leq j_x \leq L_\mathrm{NCF},j_y,\alpha,\beta}
   {(\hat{h}_{j_x}-\mu_\mathrm{NCF}\hat{\sigma}_{0})}_{\alpha,\beta}
   c_{\mathbf{j},\alpha}^\dagger
   c_{\mathbf{j},\beta},
   \\
   \hat{h}_{j_x}=&h[\hat{\sigma}_x\sin(j_x-1)\theta+\hat{\sigma}_z\cos(j_x-1)\theta].
\end{align}
Here, $H_{\mathrm{DN},l}^{L_\mathrm{NCF}}$ in Eq.~\eqref{eq:H_NCF} is given by Eq.~\eqref{eq:H_DN} by replacing $L_\mathrm{FR}$ by $L_\mathrm{NCF}$.
The schematic picture of the direction of the magnetic field is shown in Fig.~\ref{fig:Schematic_S_NCx_DN_junc}.
We also impose the periodic boundary condition in the $y$ direction with $L_y$ sites and the open-boundary condition at $j_x=L_\mathrm{NCF}+L_\mathrm{DN}$ in the $x$ direction. 

In the following, we set $\mu_\mathrm{SC}=\mu_\mathrm{FR}=\mu_\mathrm{NCF}=\mu_\mathrm{DN}=-t$~\cite{chemical_potential}, $\Delta/t=0.1$, $L_\mathrm{FR}=L_\mathrm{NCF}=5$ and $L_\mathrm{DN}=50$.

\subsection{Local density of states and Cooper pair amplitude}
In order to clarify the emergence of spin-triplet pairings and their manifestation, we calculate the local density of states (LDOS) and the anomalous Green's function.
We mainly focus on the physical quantities at the center of the DN: 
$j_x=L_\mathrm{FR}+L_\mathrm{DN}/2$ for the SC/FR/DN junction
and $j_x=L_\mathrm{NCF}+L_\mathrm{DN}/2$ for the SC/NCF/DN junction.

The Green's functions $\hat{G}_l(\tilde{z})$ of the systems are defined as
\begin{align}
   \hat{G}_l(\tilde{z}) = {(\tilde{z}-H_l)}^{-1},
\end{align}
with $\tilde{z}=E+i\eta$ (positive infinitesimal $\eta$) for the retarded Green's function and $\tilde{z}=i\omega_n$ for the Green function with Matsubara frequency [$\omega_n = (2n+1)\pi/\beta$ with inverse temperature $\beta$ and $n\in\mathbb{Z}$] representation. 
Here, the index $l$ stands for the $l$-th impurity sample.
The Green's functions are calculated by using the recursive Green's function method~\cite{PhysRevB.55.5266}.
The LDOS is obtained from the normal part of the Green's function:
\begin{align}
   \rho_\mathbf{j}(E) 
   =
   -\frac{1}{N_\mathrm{sample}\pi}\sum_{l=1}^{N_{\mathrm{sample}}}\mathrm{Im}
   \mathrm{Tr}[PG_{l,\mathbf{j},\mathbf{j}}(E+i\eta)],
\end{align}
with $\eta/t=10^{-3}$,
\begin{align}
   \hat{G}_l(\tilde{z}) =
   \begin{pmatrix}
      G_l(\tilde{z}) & F_l(\tilde{z})
      \\
      \tilde{F}_l(\tilde{z}) & \tilde{G}_l(\tilde{z})
   \end{pmatrix},
   \label{eq:LDOS}
\end{align}
and the $\mathbf{j}$-th lattice site.
Here, $P$ is a projection on the particle space: $P=(\hat{\tau}_0+\hat{\tau}_z)/2$ with a Pauli matrix $\hat{\tau}_{0,x,y,z}$ in the particle-hole space.
We discuss the averaged value of the LDOS in the $y$ direction 
\begin{align}
   \bar{\rho}_{j_x}(E) = \frac{1}{L_y}\sum_{j_y=1}^{L_y}\rho_{\mathbf{j}}(E).
\end{align}
In Sec.~\ref{sec:results}, we show the LDOS normalized by the zero energy LDOS for a normal metal (N)/FR/DN or an N/NCF/DN junction. Here, the Hamiltonian for the normal metal is $H_l$ with $\Delta=0$, and the other parameters are the same as the parameters for SC/FR/DN and SC/NCF/DN junctions, respectively.
We denote the LDOS for the N/FR/DN or the N/NCF/DN junction as $\bar{\rho}_{j_x,\mathrm{N}}(E)$.

The Cooper pair amplitude is given by the anomalous (off-diagonal) components of the Green's function $F_l(\tilde{z})$ and $\tilde{F}_l(\tilde{z})$.
Here, we focus on the component $F_l(\tilde{z})$, which has space and spin degrees of freedom:
\begin{align}
   F_{l,\mathbf{j},\mathbf{j'}}(\tilde{z})
   =&
   \begin{pmatrix}
      F_{l,\uparrow\uparrow}(\tilde{z}) & F_{l,\uparrow\downarrow}(\tilde{z})
      \\
      F_{l,\downarrow\uparrow}(\tilde{z}) & F_{l,\downarrow\downarrow}(\tilde{z})
   \end{pmatrix}_{\mathbf{j},\mathbf{j'}}
   \\
   =&
   \sum_{\alpha=0,x,y,z} f_{l,\alpha,\mathbf{j},\mathbf{j}'}(\tilde{z})\hat{\sigma}_\alpha (i\hat{\sigma}_y)
   \nonumber\\
   =&
   \begin{pmatrix}
      -f_{l,x}(\tilde{z})+if_{l,y}(\tilde{z}) & f_{l,0}(\tilde{z})+f_{l,z}(\tilde{z})
      \\
      -f_{l,0}(\tilde{z})+f_{l,z}(\tilde{z}) & f_{l,x}(\tilde{z})+if_{l,y}(\tilde{z})
   \end{pmatrix}_{\mathbf{j},\mathbf{j'}}.
\end{align}
We focus on the on-site ($\mathbf{j}=\mathbf{j}'$) component 
expressing a local $s$-wave pairing 
since it is not affected by the impurity scattering~\cite{odd1,PhysRevLett.98.107002}.
We calculate the spin-triplet $\bar{F}_{j_x,\mathrm{t},\alpha=x,y,z}(\tilde{z})$, and the spin-singlet $\bar{F}_{j_x,\mathrm{s}}(\tilde{z})$ components:
\begin{align}
   \bar{F}_{j_x,\mathrm{t},\alpha}(\tilde{z})
   =&
   \frac{1}{L_y N_\mathrm{sample}}
   \sum_{l=1}^{N_{\mathrm{sample}}}
   \sum_{j_y=1}^{L_y}
   f_{l,\alpha,\mathbf{j},\mathbf{j}}(\tilde{z}),
   \label{eq:F_t}
   \\
   \bar{F}_{j_x,\mathrm{s}}(\tilde{z})
   =&
   \frac{1}{L_y N_\mathrm{sample}}
   \sum_{l=1}^{N_{\mathrm{sample}}}
   \sum_{j_y=1}^{L_y}
   f_{l,0,\mathbf{j},\mathbf{j}}(\tilde{z}).
   \label{eq:F_s}
\end{align}
Due to the Fermi-Dirac statistics, the onsite component of the anomalous Green function 
(Cooper pair amplitude) can be categorized into two types: odd-frequency spin-triplet Cooper pair amplitude 
[$\bar{F}_{j_x,\mathrm{t},\alpha}(\tilde{z})=-\bar{F}_{j_x,\mathrm{t},\alpha}(-\tilde{z})$]
\cite{Berezinskii,LinderBalatsky,tanaka12,Cayao2020}
given by Eq.~\eqref{eq:F_t}, and even-frequency spin-singlet Cooper pair amplitude 
[$\bar{F}_{j_x,\mathrm{s}}(\tilde{z})=\bar{F}_{j_x,\mathrm{s}}(-\tilde{z})$]
given by Eq.~\eqref{eq:F_s}.

In the DN, anisotropic Cooper pairs, for example, the $p$- and $d$-wave Cooper pairs, are greatly suppressed due to impurity scattering irrespective of even or odd in frequency, and only the $s$-wave pair survives.
Then, we can concentrate on the $s$-wave Cooper pair in the DN\@. 
On the other hand, we expect that the qualitative results do not change if the DN is replaced with a ballistic metal where the polarized spin-triplet Cooper pair is also induced.
We note that for a single-band case, the even-frequency $s$-wave Cooper pair decreases the value of the LDOS at zero energy in the DN~\cite{PhysRevB.54.9443,PhysRevB.68.054513}, and the odd-frequency one increases it~\cite{odd3,odd3b,tanaka12,odd1,PhysRevB.75.134510,PhysRevLett.98.107002,PhysRevLett.98.077003}.

Also, we calculate the following quantity~\cite{RevModPhys.47.331,PhysRevB.81.014512}:
\begin{align}
   Q_{l,\alpha,\mathbf{j}}(\tilde{z})
   =&
   i{\left[\mathbf{f}_{l,\mathbf{j},\mathbf{j}}(\tilde{z}) \times \mathbf{f}_{l,\mathbf{j},\mathbf{j}}^*(\tilde{z})\right]}_\alpha,
   \label{eq:def_q}
\end{align}
with $\mathbf{f}_{l,\mathbf{j},\mathbf{j}}(\tilde{z})=(f_{l,x,\mathbf{j},\mathbf{j}}(\tilde{z}),f_{l,y,\mathbf{j},\mathbf{j}}(\tilde{z}),f_{l,z,\mathbf{j},\mathbf{j}}(\tilde{z}))$.
$Q_{l,\alpha,\mathbf{j}}(\tilde{z})$ ($\alpha=x,y,z$) is, by definition, a real quantity and expresses the polarized spin-triplet component of the Cooper pair amplitude:
\begin{align}
   Q_{l,z,\mathbf{j}}(\tilde{z})
   =&
   -\frac{1}{2}
   \left[
      {| F_{l,\uparrow\uparrow,\mathbf{j},\mathbf{j}}(\tilde{z})|}^2
      -
      {|F_{l,\downarrow\downarrow,\mathbf{j},\mathbf{j}}(\tilde{z})|}^2
   \right].
\end{align}
We discuss the averaged value of 
$\mathbf{Q}_{l,\mathbf{j}}(\tilde{z})=({Q}_{l,x,\mathbf{j}}(\tilde{z}),{Q}_{l,y,\mathbf{j}}(\tilde{z}),{Q}_{l,z,\mathbf{j}}(\tilde{z}))$:
\begin{align}
   \bar{\mathbf{Q}}_{j_x}(\tilde{z})
   =&
   \frac{1}{L_y N_\mathrm{sample}}
   \sum_{l=1}^{N_{\mathrm{sample}}}
   \sum_{j_y=1}^{L_y}
   \mathbf{Q}_{l,\mathbf{j}}(\tilde{z}).
   \label{eq:bar_q}
\end{align}
From the definition of $Q_{l,\alpha,\mathbf{j}}(\tilde{z})$ [Eq.~\eqref{eq:def_q}], $Q_{l,\alpha,\mathbf{j}}(\tilde{z})$ is the product of two odd-frequency spin-triplet Cooper pair amplitudes.
Then, $Q_{l,\alpha,\mathbf{j}}(\tilde{z})$ is an even function of frequency $\tilde{z}$.
Here, we emphasize that we use the terminology ``polarized'' spin-triplet Cooper pair for  $\bar{\mathbf{Q}}_{j_x}(\tilde{z})\neq\mathbf{0}$. 

\section{Results\label{sec:results}}
In this section, we discuss the LDOS [Eq.~\eqref{eq:LDOS}], the anomalous Green's functions [Eqs.~\eqref{eq:F_t} and \eqref{eq:F_s}], and $\bar{\mathbf{Q}}_{j_x}(i\omega_n)$ [Eq.~\eqref{eq:bar_q}].
The results for the SC/FR/DN junction are shown in Sec.~\ref{sec:result_SC_FR_DN}, and
those for the SC/NCF/DN junction are shown in Sec.~\ref{sec:result_SC_NC_DN}.
In both junctions, we obtained the polarized spin-triplet Cooper pair amplitude in the DN due to coherent spin rotation of Cooper pairs in the FR or NCF region.
\subsection{SC/FR/DN junction\label{sec:result_SC_FR_DN}}
\begin{figure}[htbp]
   \begin{center}
      \includegraphics[width=8.5cm]{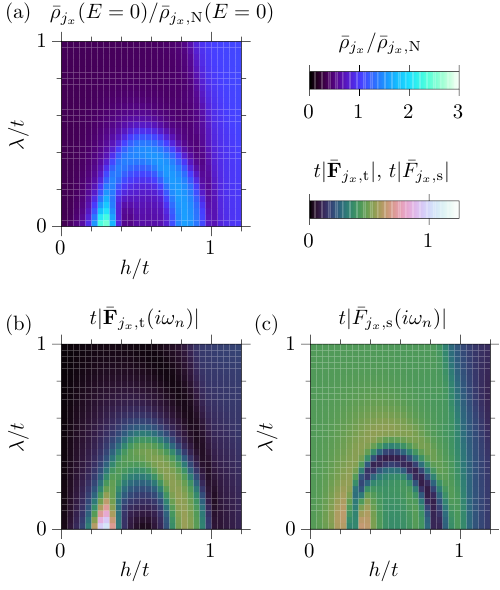}
   \end{center}
   \caption{%
      (a) The normalized LDOS $\bar{\rho}_{j_x}(E=0)/\bar{\rho}_{j_x,\mathrm{N}}(0)$,
      (b) $|\bar{\mathbf{F}}_{j_x,\mathrm{t}}(i\omega_n)|$, and
      (c) $|\bar{F}_{j_x,\mathrm{s}}(i\omega_n)|$
      are plotted as a function of $h/t$ and $\lambda/t$ at $j_x=L_\mathrm{FR}+L_\mathrm{DN}/2$.
      Here, $\bar{\rho}_{j_x,\mathrm{N}}(E)$ is the LDOS for the N/FR/DN junction.
      For the anomalous Green's functions, we fix at $\omega_n/t = 10^{-3}$.
      $L_y=100$, and $N_{\mathrm{sample}}=10^2$ samples averaged.
   }%
   \label{fig:LDOS_F_rashba}
\end{figure}
In this section, we consider the system shown in Fig.~\ref{fig:Schematic_S_FR_DN_junc}.
In Fig.~\ref{fig:LDOS_F_rashba}, we show several quantities at the center of the DN: $j_x=L_\mathrm{FR}+L_\mathrm{DN}/2$.
In Fig.~\ref{fig:LDOS_F_rashba}(a), the LDOS normalized by its normal state value is shown as a function of $h$ and $\lambda$.
The normalized LDOS exceeds unity due to the presence of the odd-frequency spin-triplet pairings 
\cite{odd3,odd3b,PhysRevB.75.134510,tanaka12,odd1}.
It is noted that the $s$-wave SC junction considered in this paper is topologically trivial, and Majorana fermions never appear at the interface.
Then this enhancement of the LDOS at zero-energy is not related to the presence of the Majorana fermion.
The normalized LDOS is the largest at approximately $(\lambda/t,h/t)=(0,0.3)$.
In Fig.~\ref{fig:LDOS_F_rashba}(b), the absolute value of the spin-triplet component $|\bar{\mathbf{F}}_{j_x,\mathrm{t}}(i\omega_n)|=\sqrt{\sum_{\alpha=x,y,z}|\bar{F}_{j_x,\mathrm{t},\alpha}(i\omega_n)|^2}$ with $\omega_n/t=10^{-3}$ is shown.
Qualitatively, $|\bar{\mathbf{F}}_{j_x,\mathrm{t}}(i\omega_n)|$ at a low frequency is very similar to the LDOS: the spin-triplet Cooper pair amplitude is generated when there is a zero-energy peak in the LDOS~\cite{odd3,odd3b,tanaka12,odd1,PhysRevB.75.134510,PhysRevLett.98.107002,PhysRevLett.98.077003}.
In Fig.~\ref{fig:LDOS_F_rashba}(c), the absolute value of the spin-singlet component $|\bar{F}_{j_x,\mathrm{s}}(i\omega_n)|$ with $\omega_n/t=10^{-3}$ is shown.
$|\bar{F}_{j_x,\mathrm{s}}(i\omega_n)|$ has a small value where the zero energy LDOS has a large value.

\begin{figure}[htbp]
   \begin{center}
      \includegraphics[width=8.5cm]{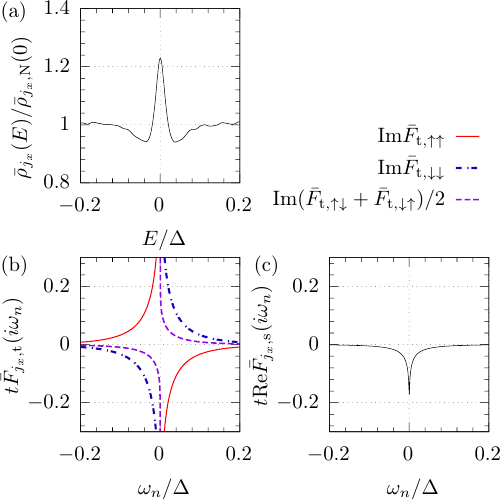}
   \end{center}
   \caption{%
      (a) The normalized LDOS is plotted as a function of $E$.
      (b) $\mathrm{Im}\bar{F}_{j_x,\mathrm{t}}(i\omega_n)$ is shown as a function of $\omega_n$. 
      (c) Re$\bar{F}_{j_x,\mathrm{s}}(i\omega_n)$ is shown as a function of $\omega_n$. 
      $\mathrm{Re}\bar{F}_{j_x,\mathrm{t},\uparrow\uparrow}(i\omega_n)=
      \mathrm{Re}\bar{F}_{j_x,\mathrm{t},\downarrow\downarrow}(i\omega_n)=
      \frac{1}{2}\mathrm{Re}[\bar{F}_{j_x,\mathrm{t},\uparrow\downarrow}(i\omega_n)+\bar{F}_{j_x,\mathrm{t},\downarrow\uparrow}(i\omega_n)]=
      \mathrm{Im}\bar{F}_{j_x,\mathrm{s}}(i\omega_n)=0$ within numerical accuracy.
      $(\lambda/t,h/t)=(0.4,0.5)$, and $j_x=L_\mathrm{FR}+L_\mathrm{DN}/2$ for (a)--(c).
      $L_y=100$, and $N_{\mathrm{sample}}=10^2$ samples averaged.
   }%
   \label{fig:energy_dep_rashba}
\end{figure}
In Fig.~\ref{fig:energy_dep_rashba}(a), the energy dependence of the normalized LDOS is shown for $(\lambda/t,h/t)=(0.4,0.5)$, where the normalized LDOS is larger than unity at zero energy.
We can see that there is a zero energy peak and corresponding to this zero energy state, the spin-triplet component of the anomalous Green's function is largely enhanced toward zero frequency [Fig.~\ref{fig:energy_dep_rashba}(b)].
The absolute value of the spin-singlet component of the anomalous Green's function also increases for low frequency, but it approaches a finite value [Fig.~\ref{fig:energy_dep_rashba}(c)].
We also calculate the normalized LDOS and the anomalous Green's function at $(\lambda/t,h/t)=(0.8,0.5)$, where we observe a gap like structure in the normalized LDOS (see Appendix~\ref{sec:App_LDOS}).

\begin{figure}[htbp]
   \begin{center}
      \includegraphics[width=8.5cm]{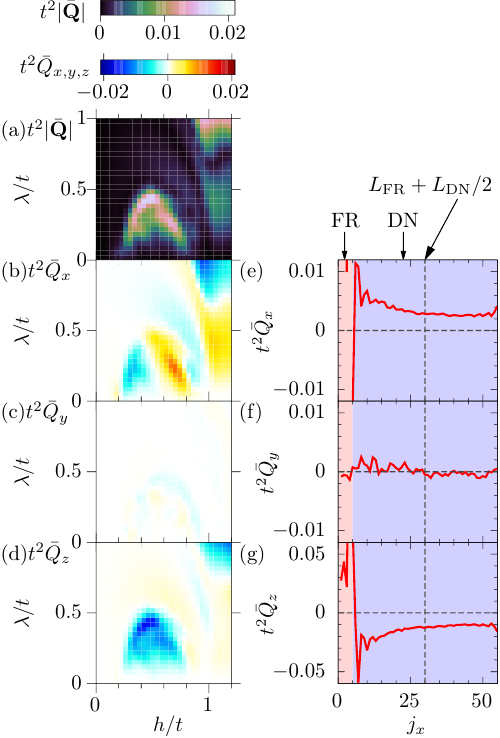}
   \end{center}
   \caption{%
      (a) $|\bar{\mathbf{Q}}_{j_x}(i\omega_n)|$,
      (b) $\bar{Q}_{j_x,x}(i\omega_n)$, 
      (c) $\bar{Q}_{j_x,y}(i\omega_n)$, and
      (d) $\bar{Q}_{j_x,z}(i\omega_n)$ are plotted as a function of $h$ and $\lambda$ at $j_x=L_\mathrm{FR}+L_\mathrm{DN}/2$.
      (e) $\bar{Q}_{j_x,x}(i\omega_n)$,
      (f) $\bar{Q}_{j_x,y}(i\omega_n)$, and
      (g) $\bar{Q}_{j_x,z}(i\omega_n)$ are plotted as a function of $j_x$ at $(\lambda/t,h/t)=(0.4,0.5)$.
      $j_x\in[1,5]$ denotes the position in the FR and,
      $j_x\in[6,55]$ denotes the position in the DN\@. 
      $L_y=100$ and $N_{\mathrm{sample}}=10^3$ samples averaged.
   }%
   \label{fig:q_SC_FR_DN}
\end{figure}

In Fig.~\ref{fig:q_SC_FR_DN}, we show $\bar{\mathbf{Q}}_{j_x}(i\omega_n)$, which reflects the polarized spin-triplet Cooper pair amplitude.
In Fig.~\ref{fig:q_SC_FR_DN}(a), we show $|\bar{\mathbf{Q}}_{j_x}(i\omega_n)|=\sqrt{\sum_{\alpha=x,y,z}\bar{Q}_{j_x,\alpha}^{2}(i\omega_n)}$. It is zero at the $h=0$ or $\lambda=0$ axis.
This means that both $h$ and $\lambda$ must be non-zero to generate the polarized spin-triplet Cooper pairing~\cite{PhysRevLett.110.117003,PhysRevB.89.134517}.
We show each component of $\bar{\mathbf{Q}}_{j_x}(i\omega_n)$ as a function of $h$ and $\lambda$ in Figs.~\ref{fig:q_SC_FR_DN}(b)--\ref{fig:q_SC_FR_DN}(d) and their spatial dependences for $(\lambda/t,h/t)=(0.4,0.5)$ in Figs.~\ref{fig:q_SC_FR_DN}(e)--\ref{fig:q_SC_FR_DN}(g).
In Figs.~\ref{fig:q_SC_FR_DN}(b)--\ref{fig:q_SC_FR_DN}(d), all the components have non zero values in some regions but 
the $y$-component is very small.
From Figs.~\ref{fig:q_SC_FR_DN}(e) and (g), we can see that the polarized spin-triplet Cooper pair amplitudes penetrate the DN\@.
Also, the $y$ component [Figs.~\ref{fig:q_SC_FR_DN}(f)] has a non zero value, but it approaches zero for a large value the of impurity sample average. 
The $N_\mathrm{sample}$ and $L_y$ dependences of the normalized LDOS, $|\bar{\mathbf{F}}_{j_x,\mathrm{t}}(i\omega_n)|$, $|\bar{F}_{j_x,\mathrm{s}}(i\omega_n)|$, and $|\bar{\mathbf{Q}}_{j_x,\mathrm{t}}(i\omega_n)|$ are discussed in Sec.~\ref{sec:app_size_FR}.
We discuss an expectation value of the spin operator to which the polarized spin-triplet Cooper pairs can contribute in Sec.~\ref{sec:app_spin_FR}.
We find that $\bar{\mathbf{Q}}_{j_x}(i\omega_n)$ and the expectation value of the spin operator are almost independent since the quasiparticles also contribute to the spin polarization.

\subsection{SC/NCF/DN junction\label{sec:result_SC_NC_DN}}

\begin{figure}[htbp]
   \begin{center}
      \includegraphics[width=8.5cm]{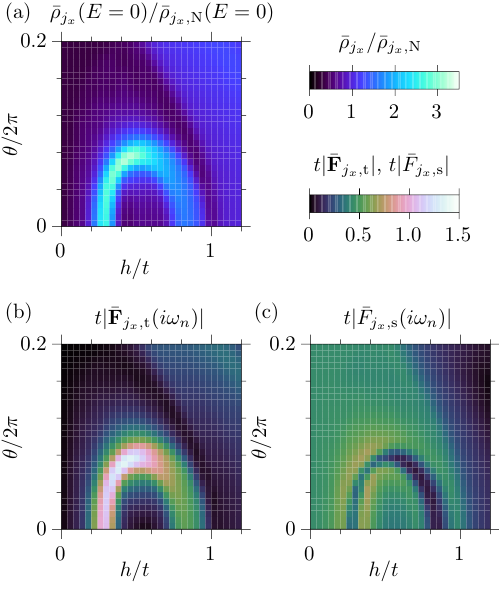}
   \end{center}
   \caption{%
      (a) The normalized LDOS at zero energy,
      (b) $|\bar{\mathbf{F}}_{j_x,\mathrm{t}}(i\omega_n)|$, and  
      (c) $|\bar{F}_{j_x,\mathrm{s}}(i\omega_n)|$
      are plotted as a function of $h$ and $\theta$ at $j_x=L_\mathrm{NCF}+L_\mathrm{DN}/2$.
      Here, $\bar{\rho}_{j_x,\mathrm{N}}(E)$ is the LDOS.
      For the anomalous Green's functions, we fix $\omega_n/t = 10^{-3}$.
      $L_y=100$, and $N_{\mathrm{sample}}=10^2$ samples averaged.
   }%
   \label{fig:LDOS_NC}
\end{figure}
Here, we show the results for the SC/NCF/DN junction.
The normalized LDOS shown in Fig.~\ref{fig:LDOS_NC}(a) exceeds unity in some regions due to the presence of the odd-frequency spin-triplet pairings [Fig.~\ref{fig:LDOS_NC}(b)]. 
The normalized LDOS for the SC/FR/DN junction is largest when $\lambda=0$, but for the SC/NCF/DN junction, the normalized LDOS is largest with non zero $\theta$.
The spin-singlet Cooper pair amplitude becomes small when the spin-triplet Cooper pair amplitude has a large value [Fig.~\ref{fig:LDOS_NC}(c)].

\begin{figure}[htbp]
   \begin{center}
      \includegraphics[width=8.5cm]{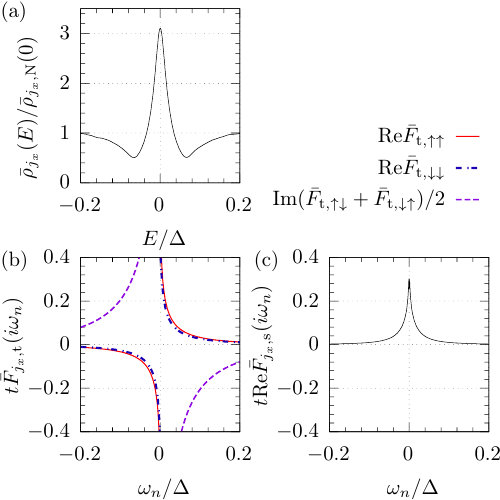}
   \end{center}
   \caption{%
      (a) The normalized LDOS is plotted as a function of $E$.
      (b) $\bar{F}_{j_x,\mathrm{t}}(i\omega_n)$ is shown as a function of $\omega_n$. 
      (c) Re$\bar{F}_{j_x,\mathrm{s}}(i\omega_n)$ is shown as a function of $\omega_n$. 
      $\mathrm{Im}\bar{F}_{j_x,\mathrm{t},\uparrow\uparrow}(i\omega_n)=
      \mathrm{Im}\bar{F}_{j_x,\mathrm{t},\downarrow\downarrow}(i\omega_n)=
      \mathrm{Re}[\bar{F}_{j_x,\mathrm{t},\uparrow\downarrow}(i\omega_n)+\bar{F}_{j_x,\mathrm{t},\downarrow\uparrow}(i\omega_n)]=
      \mathrm{Im}\bar{F}_{j_x,\mathrm{s}}(i\omega_n)=0$ within numerical accuracy.
      $(\theta/2\pi,h/t)=(0,075,0.5)$, and $j_x=L_\mathrm{NCF}+L_\mathrm{DN}/2$ for (a)--(c).
      $L_y=100$, and $N_{\mathrm{sample}}=10^2$ samples averaged.
   }%
   \label{fig:energy_dep_NC}
\end{figure}
In Fig.~\ref{fig:energy_dep_NC}(a), we show the energy dependence of the normalized LDOS at $(\theta/2\pi,h/t)=(0.075,0.5)$ 
and see that it has a zero energy peak.
In Figs.~\ref{fig:energy_dep_NC}(b) and \ref{fig:energy_dep_NC}(c), the frequency dependences of the spin-triplet and the spin-singlet Cooper pair amplitudes are shown, respectively.
The spin-triplet Cooper pair amplitude is also largely enhanced toward zero frequency, and the spin-singlet one has a non zero value for $\omega_n\rightarrow 0$.
We also show the normalized LDOS and the anomalous Green's  function at $(\theta/2\pi,h/t)=(0.15,0.5)$ in Appendix~\ref{sec:App_LDOS}.

\begin{figure}[htbp]
   \begin{center}
      \includegraphics[width=8.5cm]{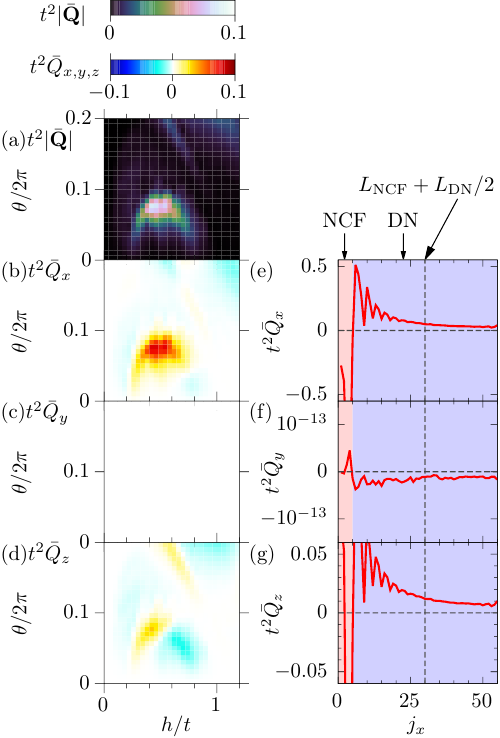}
   \end{center}
   \caption{%
      (a) $|\bar{\mathbf{Q}}_{j_x}(i\omega_n)|$,
      (b) $\bar{Q}_{j_x,x}(i\omega_n)$, 
      (c) $\bar{Q}_{j_x,y}(i\omega_n)$, and
      (d) $\bar{Q}_{j_x,z}(i\omega_n)$ are plotted as a function of $h$ and $\theta$ for $j_x=L_\mathrm{NCF}+L_\mathrm{DN}/2$.
      (e) $\bar{Q}_{j_x,x}(i\omega_n)$,
      (f) $\bar{Q}_{j_x,y}(i\omega_n)$, and
      (g) $\bar{Q}_{j_x,z}(i\omega_n)$ are plotted as a function of $j_x$.
      $j_x\in[1,5]$ is the position in the NCF and,
      $j_x\in[6,55]$ is the position in the DN. 
      In (e)--(g) $(\theta/2\pi,h/t)=(0.08,0.5)$.
      $L_y=100$ and $N_{\mathrm{sample}}=10^3$ samples averaged.
   }
   \label{fig:q_SC_NC_DN}
\end{figure}
The polarized spin-triplet Cooper pair amplitudes are generated in the SC/NCF/DN junction (Fig.~\ref{fig:q_SC_NC_DN}).
In Fig.~\ref{fig:q_SC_NC_DN}(a), the absolute value of $\bar{\mathbf{Q}}_{j_x}(i\omega_n)$, with $\omega_n/t=10^{-3}$, is shown,
and it is zero for $\theta=0$ or $h=0$.
The $h=0$ case is trivial: there is no field, and the Hamiltonian has spin rotational symmetry.
Here, the polarized spin-triplet Cooper pair amplitude has only $x$ and $z$ components [Figs.~\ref{fig:q_SC_NC_DN}(b)--\ref{fig:q_SC_NC_DN}(d)], and it might correspond to the fact that
the noncollinear spin structure in the NCF lies in the $x-z$ plane~\cite{PhysRevB.71.220506,PhysRevB.72.054523}.
We also show the $j_x$ dependence of $\bar{\mathbf{Q}}_{j_x}(i\omega_n)$ in Figs.~\ref{fig:q_SC_NC_DN}(e)--\ref{fig:q_SC_NC_DN}(g).
Similar to the SC/FR/DN junction, the polarized spin-triplet Cooper pair amplitudes penetrate the DN\@.
The $y$ component of $\bar{\mathbf{Q}}_{j_x}(i\omega_n)$ is almost zero [Figs.~\ref{fig:q_SC_NC_DN}(f)].
The dependences of these results on $N_\mathrm{sample}$ and $L_y$ are discussed in Sec.~\ref{sec:app_size_NC}.
We also discuss the expectation value of the spin operator in Sec.~\ref{sec:app_spin_NC}, and we do not find a direct relationship between the expectation value of the spin operator and $\bar{\mathbf{Q}}_{j_x}(i\omega_n)$.

\section{Summary\label{sec:summary}}
In this paper, we showed that the polarized spin-triplet Cooper pair amplitude is generated by the spin-singlet $s$-wave superconductor in two kinds of junctions: the $s$-wave SC/ferromagnetic metal with Rashba spin-orbit coupling/diffusive normal metal junction and the $s$-wave SC/noncollinear ferromagnetic metal/diffusive normal metal junction.
We have clarified the generation of the spin polarized triplet pairings in the diffusive normal metal due to coherent spin rotation in the magnetic regions. The emergence of the triplet pairings manifests as a zero-energy peak in the density of states.
 Candidate magnets for these junctions are magnets without inversion symmetry\cite{https://doi.org/10.1002/adma.201603227,doi:10.1063/1.5139488} such as MnSi\cite{PhysRevLett.102.197202}, MnGe~\cite{PhysRevLett.106.156603}, (V,Pt)Se$_2$~\cite{Velez-Fort2022}, and NbMnP~\cite{PhysRevB.104.174413}. Although we have performed numerical calculations in two-dimensional systems due to computational cost, we expect that we can obtain qualitatively the same results for three-dimensional junctions with a magnetic interface.

In this paper, we have chosen the spin-singlet $s$-wave pairing as the symmetry of the Cooper pair in the SC. 
If we choose spin-singlet $d$-wave 
pairing, zero-energy Andreev bound states~\cite{Hu94,TK95,Kashiwaya_2000}
and the resulting odd-frequency pairing~\cite{tanaka12}
protected by the spectral 
bulk-edge correspondence are
generated at the interface~\cite{Spectralbulk}. 
It is an interesting issue to study 
the proximity effect in $d$-wave 
superconductor junctions~\cite{TamuraProximity}
in the presence of a ferromagnetic metal with Rashba spin-orbit coupling or a noncollinear ferromagnetic metal. 
Also, although we concentrated on superconductor junctions with a magnetic interface as the source of the polarized spin-triplet Cooper pair in this paper, a combination of the $s$-wave SC and topological systems is also interesting since many topological systems require spin-orbit interaction or a magnetic field such as nanowire systems~\cite{Asano2013,PhysRevB.95.184506,PhysRevB.98.075425} and SC/topological insulator junctions~\cite{PhysRevB.86.075410,PhysRevB.86.144506,PhysRevB.87.220506,PhysRevB.92.100507,PhysRevB.92.205424,Vasenko_2017,PhysRevB.96.155426,PhysRevLett.125.026802}.

\begin{acknowledgments}
This work was supported by 
JSPS KAKENHI Grants 
No. JP20H00131, No. JP18H01176, No. JP20H01857,  and 
No. JP30578216 and the JSPS-EPSRC Core-to-Core program ``Oxide Superspin.''
\end{acknowledgments}

\appendix

\section{Energy or frequency dependence of LDOS and anomalous Green's function\label{sec:App_LDOS}}
\begin{figure}[htbp]
   \begin{center}
      \includegraphics[width=8.5cm]{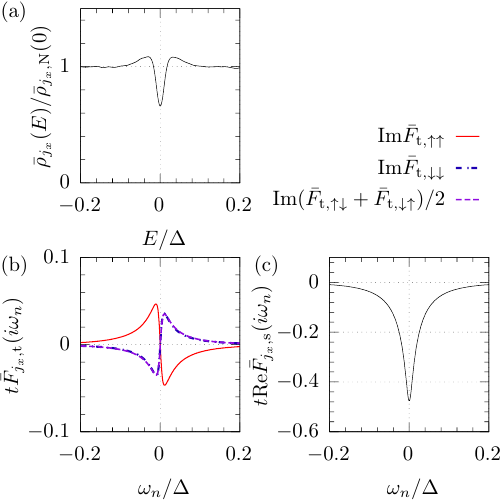}
   \end{center}
   \caption{%
      (a) The normalized LDOS is plotted as a function of $E$.
      (b) $\mathrm{Im}\bar{F}_{j_x,\mathrm{t}}(i\omega_n)$ is shown as a function of $\omega_n$. 
      (c) Re$\bar{F}_{j_x,\mathrm{s}}(i\omega_n)$ is shown as a function of $\omega_n$. 
      $\mathrm{Re}\bar{F}_{j_x,\mathrm{t},\uparrow\uparrow}(i\omega_n)=
      \mathrm{Re}\bar{F}_{j_x,\mathrm{t},\downarrow\downarrow}(i\omega_n)=
      \frac{1}{2}\mathrm{Re}[\bar{F}_{j_x,\mathrm{t},\uparrow\downarrow}(i\omega_n)+\bar{F}_{j_x,\mathrm{t},\downarrow\uparrow}(i\omega_n)]=
      \mathrm{Im}\bar{F}_{j_x,\mathrm{s}}(i\omega_n)=0$ within numerical accuracy.
      $(\lambda/t,h/t)=(0.8,0.5)$, $j_x=L_\mathrm{FR}+L_\mathrm{DN}/2$,
      $L_y=100$, and $N_{\mathrm{sample}}=10^2$ samples averaged.
   }%
   \label{fig:energy_dep_rashba_app}
\end{figure}

\begin{figure}[htbp]
   \begin{center}
      \includegraphics[width=8.5cm]{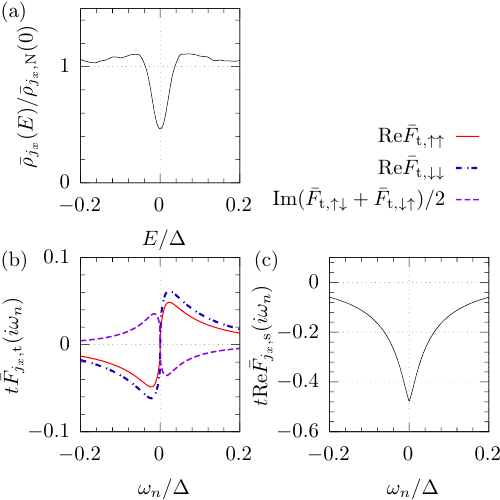}
   \end{center}
   \caption{%
      (a) The normalized LDOS is plotted as a function of $E$
      (b) $\bar{F}_{j_x,\mathrm{t}}(i\omega_n)$ is shown as a function of $\omega_n$. 
      (c) Re$\bar{F}_{j_x,\mathrm{s}}(i\omega_n)$ is shown as a function of $\omega_n$. 
      $\mathrm{Im}\bar{F}_{j_x,\mathrm{t},\uparrow\uparrow}(i\omega_n)=
      \mathrm{Im}\bar{F}_{j_x,\mathrm{t},\downarrow\downarrow}(i\omega_n)=
      \mathrm{Re}[\bar{F}_{j_x,\mathrm{t},\uparrow\downarrow}(i\omega_n)+\bar{F}_{j_x,\mathrm{t},\downarrow\uparrow}(i\omega_n)]=
      \mathrm{Im}\bar{F}_{j_x,\mathrm{s}}(i\omega_n)=0$ within numerical accuracy.
      $(\theta/2\pi,h/t)=(0.15,0.5)$, $j_x=L_\mathrm{NCF}+L_\mathrm{DN}/2$,
      $L_y=100$ and $N_{\mathrm{sample}}=10^2$ samples averaged.
   }%
   \label{fig:energy_dep_NC_app}
\end{figure}

In Fig.~\ref{fig:energy_dep_rashba_app} (SC/FR/DN junction) and Fig.~\ref{fig:energy_dep_NC_app} (SC/NCF/DN junction), we show the energy dependence of the normalized LDOS and
$\omega_n$ dependences of $\bar{\mathbf{F}}_{j_x,\mathrm{t}}(i\omega_n)$, and $\bar{F}_{j_x,\mathrm{s}}(i\omega_n)$
at the parameter where the normalized LDOS at zero energy is smaller than unity.
In both graphs, we can see that there is a gap like structure at zero energy for the normalized LDOS 
[Figs.~\ref{fig:energy_dep_rashba_app}(a) and \ref{fig:energy_dep_NC_app}(a)].
The spin-triplet and spin-singlet components of the anomalous Green's function are shown in
Figs.~\ref{fig:energy_dep_rashba_app}(b), \ref{fig:energy_dep_rashba_app}(c) and \ref{fig:energy_dep_NC_app}(b), \ref{fig:energy_dep_NC_app}(c).
The absolute value of the spin-singlet component is larger than that of each spin-triplet component, and the spin-triplet components are linear functions at $\omega_n=0$.


\section{System size and sample number dependence\label{sec:App_size}}
Here we show the $N_\mathrm{sample}$ and $L_y$ dependences of the normalized LDOS, 
$|\bar{\mathbf{F}}_{j_x,\mathrm{t}}(i\omega_n)|$, $|\bar{F}_{j_x,\mathrm{s}}(i\omega_n)|$, and $|\bar{\mathbf{Q}}_{j_x}(i\omega_n)|$ in order to demonstrate the robustness of our results.
\subsection{SC/FR/DN junction\label{sec:app_size_FR}}
\begin{figure}[htbp]
    \centering
    \includegraphics[width=8.5cm]{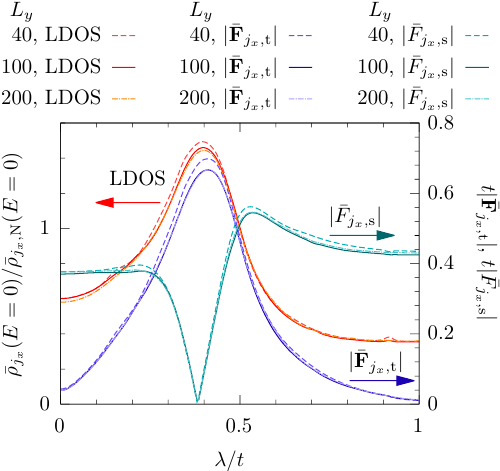}
    \caption{%
    The normalized zero energy LDOS, $|\bar{\mathbf{F}}_{j_x,\mathrm{t}}(i\omega_n)|$, and $|\bar{F}_{j_x,\mathrm{s}}(i\omega_n)|$
    are plotted as a function of $\lambda$ with $h/t=0.5$ and $\omega_n/t=10^{-3}$ 
    for several values of $L_y$.
    $j_x=L_{\mathrm{FR}}+L_\mathrm{DN}/2$, and
    $N_{\mathrm{sample}}=10^2$ samples averaged.
    }
    \label{fig:size_SC_FR_DN}
\end{figure}

\begin{figure}[htbp]
    \centering
    \includegraphics[width=8.5cm]{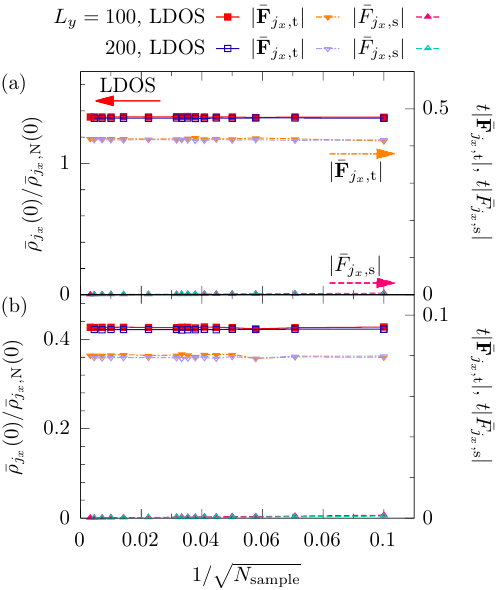}
    \caption{%
    The normalized zero-energy LDOS, $|\bar{\mathbf{F}}_{j_x,\mathrm{t}}(i\omega_n)|$, and $|\bar{F}_{j_x,\mathrm{s}}(i\omega_n)|$ with $\omega_n/t=10^{-3}$ are plotted as a function of $1/\sqrt{N_\mathrm{sample}}$
    for (a) $(\lambda/t,h/t)=(0.35,0.5)$ and (b) $(0.7,0.5)$.
    $j_x=L_{\mathrm{FR}}+L_\mathrm{DN}/2$
    with $L_y=100$ and $200$.
    }
    \label{fig:N_dos_SC_FR_DN}
\end{figure}

\begin{figure}[htbp]
    \centering
    \includegraphics[width=8.5cm]{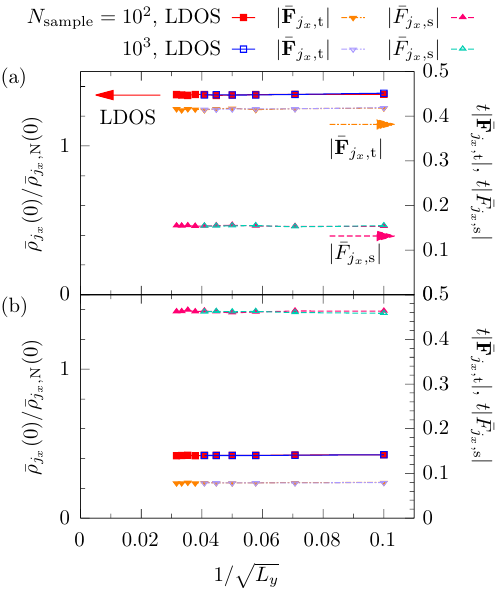}
    \caption{%
    The normalized zero-energy LDOS, $|\bar{\mathbf{F}}_{j_x,\mathrm{t}}(i\omega_n)|$, and $|\bar{F}_{j_x,\mathrm{s}}(i\omega_n)|$ with $\omega_n/t=10^{-3}$ are plotted as a function of $1/\sqrt{L_y}$
    for (a) $(\lambda/t,h/t)=(0.35,0.5)$ and (b) $(0.7,0.5)$.
    $j_x=L_{\mathrm{FR}}+L_\mathrm{DN}/2$
    with $N_\mathrm{sample}=10^2$ and $10^3$.
    }
    \label{fig:L_y_dos_SC_FR_DN}
\end{figure}

We show the $L_y$ dependences of the normalized LDOS at zero energy, $|\bar{\mathbf{F}}_{j_x,\mathrm{t}}(i\omega_n)|$, and $|\bar{F}_{j_x,\mathrm{s}}(i\omega_n)|$ at a low frequency
in Fig.~\ref{fig:size_SC_FR_DN}.
Their system size dependences are not significant.
In Figs.~\ref{fig:N_dos_SC_FR_DN} and \ref{fig:L_y_dos_SC_FR_DN}, we show $N_\mathrm{sample}$
and $L_y$ dependences, respectively, at fixed $\lambda$ and $h$.
The $N_\mathrm{sample}$ and $L_y$ dependences are also not significant in these plots.

\begin{figure}[htbp]
    \centering
    \includegraphics[width=8.5cm]{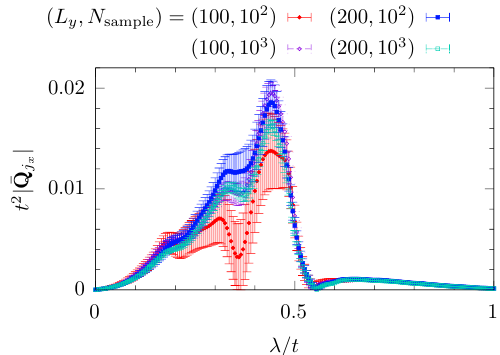}
    \caption{%
    $|\bar{\mathbf{Q}}_{j_x}(i\omega_n)|$ is plotted as a function of $\lambda/t$ for $L_y=100$ and $200$, and $N_\mathrm{sample}=10^2$ and $10^3$ with $h/t=0.5$ and $\omega_n/t=10^{-3}$.
    $j_x=L_{\mathrm{FR}}+L_\mathrm{DN}/2$.
    }
    \label{fig:q_sample_SC_FR_DN}
\end{figure}

\begin{figure}[htbp]
    \centering
    \includegraphics[width=8.5cm]{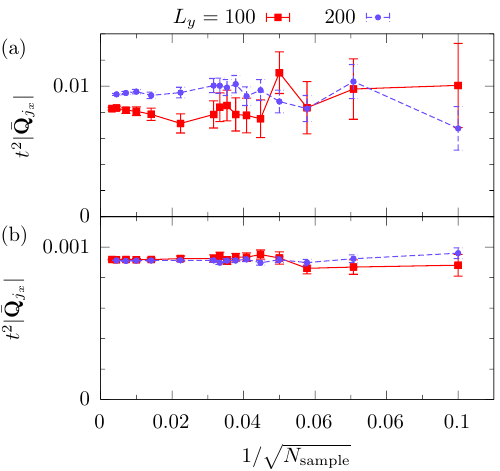}
    \caption{%
    $|\bar{\mathbf{Q}}_{j_x}(i\omega_n)|$ is plotted as a function of $1/\sqrt{N_\mathrm{sample}}$
    for (a) $\lambda/t=0.35$,  and (b) $0.7$ with $h/t=0.5$ and $\omega_n/t=10^{-3}$.
    $j_x=L_{\mathrm{FR}}+L_\mathrm{DN}/2$.
    }
    \label{fig:sample_SC_FR_DN}
\end{figure}

\begin{figure}[htbp]
    \centering
    \includegraphics[width=8.5cm]{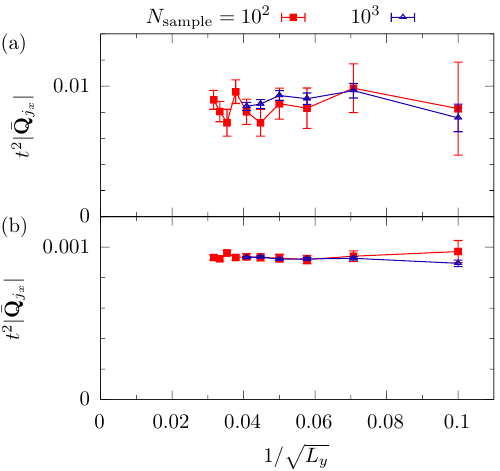}
    \caption{%
    $|\bar{\mathbf{Q}}_{j_x}(i\omega_n)|$ is plotted as a function of $1/\sqrt{L_y}$
    for (a) $(\lambda/t,h/t)=(0.35,0.5)$, and (b) $(0.7,0.5)$ with $\omega_n/t=10^{-3}$.
    $j_x=L_{\mathrm{FR}}+L_\mathrm{DN}/2$.
    }
    \label{fig:L_y_q_SC_FR_DN}
\end{figure}

In Fig.~\ref{fig:q_sample_SC_FR_DN}, we show $|\bar{\mathbf{Q}}_{j_x}(i\omega_n)|$ for $(L_y,N_\mathrm{sample})=(100,10^2)$, $(100,10^3)$, $(200,10^2)$, and $(200,10^3)$ at $\omega_n/t=10^{-3}$.
Here, we show the standard error as an error bar.
The average values $|\bar{\mathbf{Q}}_{j_x}(i\omega_n)|$ have a large fluctuation for $0.1\lesssim\lambda/t\lesssim 0.5$. 
It might correspond to the fact that the normalized LDOS at zero energy has large values close to these parameters.
For $N_\mathrm{sample}=\mathrm{10^3}$, $|\bar{\mathbf{Q}}_{j_x}(i\omega_n)|$ for $L_y=100$ and $L_y=200$ have almost the same value.
In Fig.~\ref{fig:sample_SC_FR_DN}, we show the $N_\mathrm{sample}$ dependence of $|\bar{\mathbf{Q}}_{j_x}(i\omega_n)|$ for $h/t=0.5$.
$|\bar{\mathbf{Q}}_{j_x}(i\omega_n)|$ for $L_y=100$ and $200$ have almost the same value but for large $N_\mathrm{sample}$, $|\bar{\mathbf{Q}}_{j_x}(i\omega_n)|$ with $L_y=100$ has a slightly smaller value at $\lambda/t=0.35$ [Fig.~\ref{fig:sample_SC_FR_DN}(a)].
In Fig.~\ref{fig:L_y_q_SC_FR_DN}, we show the $L_y$ dependence of $|\bar{\mathbf{Q}}_{j_x}(i\omega_n)|$ at $h/t=0.5$ for $N_\mathrm{sample}=10^2$ and $10^3$.
There are somewhat large statistical errors for $N_\mathrm{sample}=10^2$, but the size of the error bar is sufficiently small for $L_y=100$ and $N_\mathrm{sample}=10^3$.


\subsection{SC/NCF/DN junction\label{sec:app_size_NC}}
\begin{figure}[htbp]
    \centering
    \includegraphics[width=8.5cm]{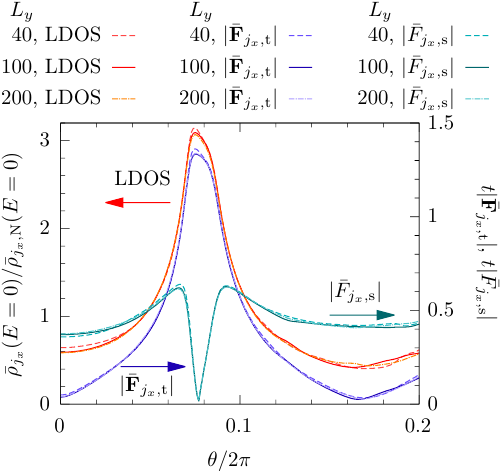}
    \caption{%
    The normalized LDOS, $|\bar{\mathbf{F}}_{j_x,\mathrm{t}}(i\omega_n)|$, and $|\bar{F}_{j_x,\mathrm{s}}(i\omega_n)|$
    are plotted as a function of $\theta$ with $h/t=0.5$ and $\omega_n/t=10^{-3}$ 
    for several values of $L_y$.
    $j_x=L_{\mathrm{NCF}}+L_\mathrm{DN}/2$ and
    $N_{\mathrm{sample}}=10^2$ sample averaged.
    }
    \label{fig:size_SC_NC_DN}
\end{figure}

\begin{figure}[htbp]
    \centering
    \includegraphics[width=8.5cm]{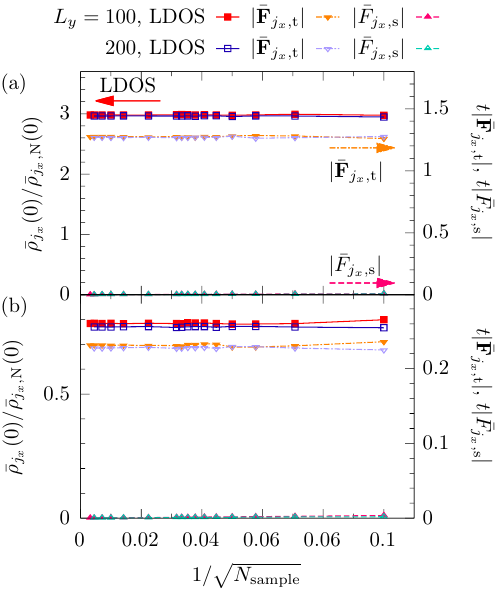}
    \caption{%
    The normalized zero energy LDOS, $|\bar{\mathbf{F}}_{j_x,\mathrm{t}}(i\omega_n)|$, and $|\bar{F}_{j_x,\mathrm{s}}(i\omega_n)|$ with $\omega_n/t=10^{-3}$ are plotted as a function of $1/\sqrt{N_\mathrm{sample}}$
    for (a) $(\theta/2\pi,h/t)=(0.08,0.5)$ and (b) $(0.12,0.5)$.
    $j_x=L_{\mathrm{NCF}}+L_\mathrm{DN}/2$.
    }
    \label{fig:N_dos_SC_NC_DN}
\end{figure}

\begin{figure}[htbp]
    \centering
    \includegraphics[width=8.5cm]{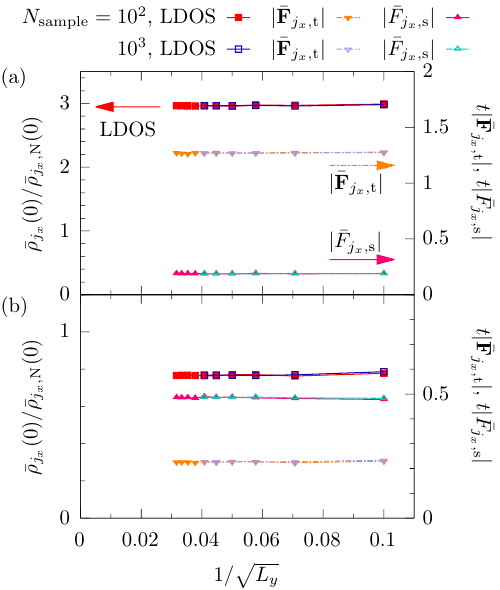}
    \caption{%
    The normalized zero energy LDOS, $|\bar{\mathbf{F}}_{j_x,\mathrm{t}}(i\omega_n)|$, and $|\bar{F}_{j_x,\mathrm{s}}(i\omega_n)|$ with $\omega_n/t=10^{-3}$ are plotted as a function of $1/\sqrt{L_y}$
    for (a) $\theta/2\pi=0.08$ and (b) $0.12$ with $h/t=0.5$.
    $j_x=L_{\mathrm{NCF}}+L_\mathrm{DN}/2$.
    }
    \label{fig:L_y_dos_SC_NC_DN}
\end{figure}

We show the $N_\mathrm{sample}$ and $L_y$ dependences for the SC/NCF/DN junction.
We show the $L_y$ dependence of the normalized LDOS, $|\bar{\mathbf{F}}_{j_x,\mathrm{t}}(i\omega_n)|$, and $|\bar{F}_{j_x,\mathrm{s}}(i\omega_n)|$ at a low frequency
in Fig.~\ref{fig:size_SC_NC_DN}.
In this case, similar to the SC/FR/DN junction,
their system size dependences are weak.
In Figs.~\ref{fig:N_dos_SC_NC_DN} and \ref{fig:L_y_dos_SC_NC_DN}, we show the $N_\mathrm{sample}$
and $L_y$ dependences, respectively, at fixed $\theta$ and $h$.
The $N_\mathrm{sample}$ and $L_y$ dependences are also not very strong in these plots.

\begin{figure}[htbp]
    \centering
    \includegraphics[width=8.5cm]{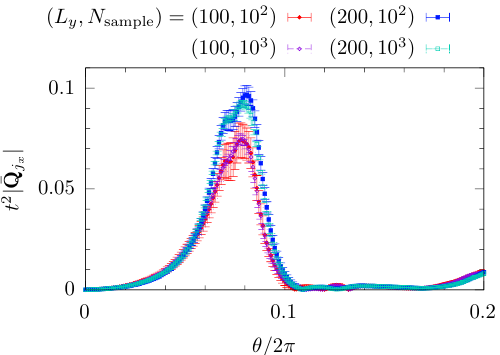}
    \caption{%
    $|\bar{\mathbf{Q}}_{j_x}(i\omega_n)|$ is plotted as a function of $\theta$ for several $L_y$ and $N_\mathrm{sample}$ with $h/t=0.5$ and $\omega_n/t=10^{-3}$.
    $j_x=L_{\mathrm{NCF}}+L_\mathrm{DN}/2$.
    }
    \label{fig:theta_dep_q_SC_NC_DN}
\end{figure}

\begin{figure}[htbp]
    \centering
    \includegraphics[width=8.5cm]{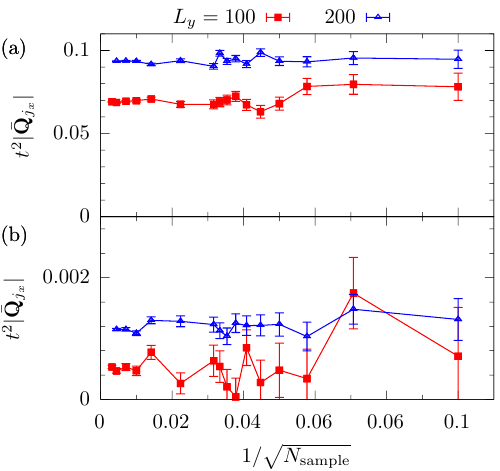}
    \caption{%
    $|\bar{\mathbf{Q}}_{j_x}(i\omega_n)|$ is plotted as a function of $1/\sqrt{N_\mathrm{sample}}$
    for (a) $\theta/2\pi=0.08$, and (b) $0.12$ with $h/t=0.5$, and $\omega_n/t=10^{-3}$.
    $j_x=L_{\mathrm{NCF}}+L_\mathrm{DN}/2$.
    }
    \label{fig:N_F_F_SC_NC_DN}
\end{figure}

\begin{figure}[htbp]
    \centering
    \includegraphics[width=8.5cm]{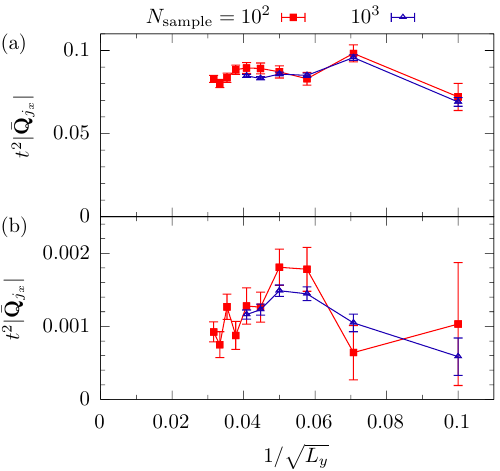}
    \caption{%
    $|\bar{\mathbf{Q}}_{j_x}(i\omega_n)|$ is plotted as a function of $1/\sqrt{L_y}$
    for (a) $\theta/2\pi=0.08$, and (b) $0.12$ with $h/t=0.5$ and $\omega_n/t=10^{-3}$.
    $j_x=L_{\mathrm{NCF}}+L_\mathrm{DN}/2$.
    }
    \label{fig:L_y_q_SC_NC_DN}
\end{figure}

In Fig.~\ref{fig:theta_dep_q_SC_NC_DN}, we show the $|\bar{\mathbf{Q}}_{j_x}(i\omega_n)|$ for $(L_y,N_\mathrm{sample})=(100,10^2)$, $(100,10^3)$, $(200,10^2)$, and $(200,10^3)$ at $\omega_n/t=10^{-3}$.
Here, we also show the standard error as an error bar.
We also observe a large error at $0.06\lesssim\theta/2\pi\lesssim 0.09$. 
$|\bar{\mathbf{Q}}_{j_x}(i\omega_n)|$ for $N_\mathrm{sample}=10^2$ and $10^3$ have almost the same value but $L_y=100$ and $200$ are different for $0.06\lesssim\theta/2\pi\lesssim 0.09$.
In Fig.~\ref{fig:N_F_F_SC_NC_DN}, we show the $N_\mathrm{sample}$ dependence of $|\bar{\mathbf{Q}}_{j_x}(i\omega_n)|$ for $h/t=0.5$.
$|\bar{\mathbf{Q}}_{j_x}(i\omega_n)|$ for $L_y=100$ is smaller than that for $200$ even for large $N_\mathrm{sample}$, at least for $\theta/2\pi=0.08$ [Fig.~\ref{fig:N_F_F_SC_NC_DN}(a)].
In Fig.~\ref{fig:L_y_q_SC_NC_DN}, we show the $L_y$ dependence of $|\bar{\mathbf{Q}}_{j_x}(i\omega_n)|$ at $(\theta/2\pi,h/t)=(0.08,0.5)$ and $(0.12,0,5)$ for $N_\mathrm{sample}=10^2$ and $10^3$.
The $L_y$ dependence is not small for $\theta/2\pi=0.08$ [Fig.~\ref{fig:L_y_q_SC_NC_DN}(a)] but we expect that the qualitative behaviors of the results do not change.


\section{Expectation value of spin operator\label{sec:App_spin}}
In this appendix, we discuss the expectation of the spin operator:
\begin{align}
    \bar{s}_{j_x,\alpha} = \frac{1}{L_y N_\mathrm{sample}\beta}\sum_{j_y=1}^{L_y}\sum_{l=1}^{N_\mathrm{sample}}\sum_n \mathrm{Tr}[P\hat{\sigma}_\alpha G_{l,\mathbf{j},\mathbf{j}}(i\omega_n)],
\end{align}
with $\alpha=x,y,z$ and the inverse of the temperature $\beta$.
Here, we use an IR basis to calculate the Matsubara frequency sum~\cite{CHIKANO2019181} which reduces 
the computational cost.
In the following, we choose $\beta t=10^3$. 
We confirmed that the results are almost the same as the results with $\beta t =10^4$, and thus, $\beta t =10^3$ is a sufficiently small temperature.
\subsection{SC/FR/DN junction\label{sec:app_spin_FR}}
\begin{figure}[htbp]
    \centering
    \vspace{5mm}
    \includegraphics[width=8.5cm]{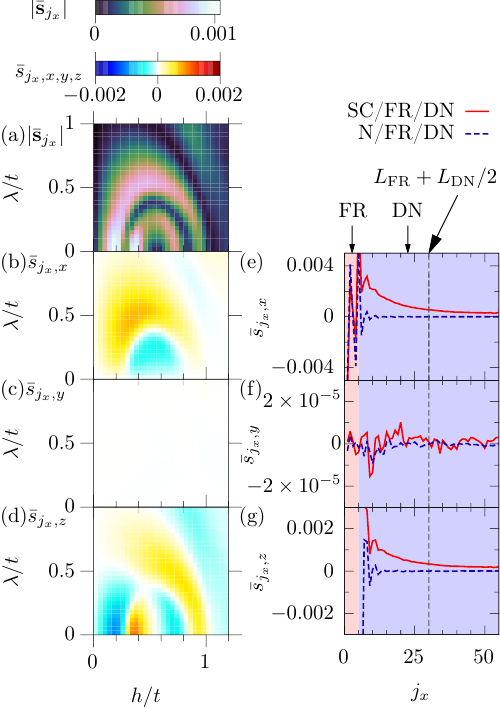}
    \caption{%
    (a) $|\bar{\mathbf{s}}_{j_x}|$,
    (b) $\bar{s}_{j_x,x}$,
    (c) $\bar{s}_{j_x,y}$, and
    (d) $\bar{s}_{j_x,z}$,
    are plotted as a function of $h$ and $\lambda$ at $j_x=L_\mathrm{FR}+L_\mathrm{DN}/2$.
    (e) $\bar{s}_{j_x,x}$,
    (f) $\bar{s}_{j_x,y}$, and
    (g) $\bar{s}_{j_x,z}$ are plotted as a function of $j_x$ for $(\lambda/t,h/t)=(0.5,0.5)$.
    $t\beta=10^3$, $L_y=40$, and $N_{\mathrm{sample}}=10^2$ samples averaged.
    }
    \label{fig:spin_SC_FR_DN}
\end{figure}

\begin{figure}[htbp]
    \centering
    \includegraphics[width=8.5cm]{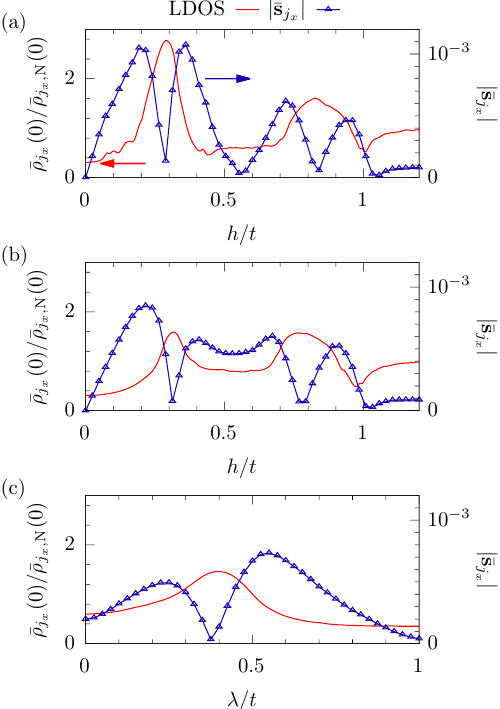}
    \caption{%
    The normalized zero energy LDOS (left $y$-axis) and $|\bar{\mathbf{s}}_{j_x}|$ (right $y$-axis) are plotted as a function of 
    $h/t$ for (a) $\lambda=0$, and
    (b) $\lambda/t=0.2$ and plotted as a function of $\lambda/t$ for
    (c) $h/t=0.5$ at $j_x=L_\mathrm{FR}+L_\mathrm{DN}/2$.
    $t\beta=10^3$, $L_y=100$, and $N_{\mathrm{sample}}=10^2$ samples averaged.
    }
    \label{fig:spin_cut_SC_FR_DN}
\end{figure}

In Fig.~\ref{fig:spin_SC_FR_DN}, we show the expectation value of the spin operator [Figs.~\ref{fig:spin_SC_FR_DN}(a)--\ref{fig:spin_SC_FR_DN}(d)] and its spatial dependence [Figs.~\ref{fig:spin_SC_FR_DN}(e)--\ref{fig:spin_SC_FR_DN}(g)] for the SC/FR/DN junction. 
$|\bar{\mathbf{s}}_{j_x}|$ has a small value in the region where the normalized LDOS at zero energy exceeds unity and  has a large value [Fig.~\ref{fig:LDOS_F_rashba}(a), see also Fig.~\ref{fig:spin_cut_SC_FR_DN}].
This might be understood as follows: the LDOS at zero energy might have a maximum when both the spin-up and -down components of the LDOS have large values, and as a consequence, the expectation value of the spin becomes small (minimum) in the DN. 

$|\bar{\mathbf{s}}_{j_x}|$ becomes non zero at $\lambda=0$ in some region where the polarized spin-triplet 
Cooper pair amplitude is zero.
In general, both the quasiparticles and the Cooper pairs contribute to $\bar{\mathbf{s}}_{j_x}$.
Thus, the non zero value of $\bar{\mathbf{s}}_{j_x}$ at $\lambda=0$ comes from spin polarization of the quasiparticles.
Therefore, the non zero value of $\bar{\mathbf{s}}_{j_x}$ is not direct evidence of the presence 
of the polarized spin-triplet Cooper pair amplitude.
$\bar{\mathbf{s}}_{j_x}$ has only $x$ and $z$ components similar to $\bar{\mathbf{Q}}_{j_x}(i\omega_n)$ in Figs.~\ref{fig:q_SC_FR_DN}(b)--\ref{fig:q_SC_FR_DN}(d).
In Figs.~\ref{fig:spin_SC_FR_DN}(e)--\ref{fig:spin_SC_FR_DN}(g), we show the spatial dependence of $\bar{\mathbf{s}}_{j_x}$ for
the SC/FR/DN and N/FR/DN junctions (for N, we set the pair potential $\Delta=0$, and the other parameters are the same as in the corresponding SC junctions).
We can see that the presence of the SC is crucial for the non zero value of $\bar{s}_{j_x,x}$, and
$\bar{s}_{j_x,z}$ in the DN.

\subsection{SC/NCF/DN junction\label{sec:app_spin_NC}}
\begin{figure}[htbp]
    \centering
    \includegraphics[width=8.5cm]{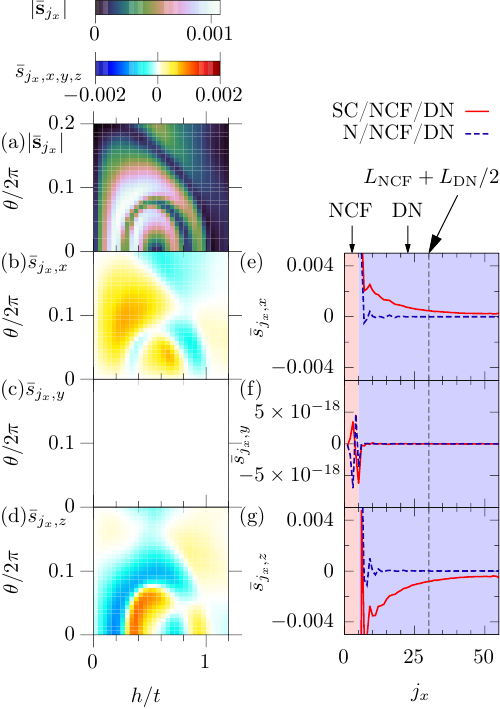}
    \caption{%
    (a) $|\bar{\mathbf{s}}_{j_x}|$,
    (b) $\bar{s}_{j_x,x}$,
    (c) $\bar{s}_{j_x,y}$, and
    (d )$\bar{s}_{j_x,z}$,
    are plotted as a function of $h$ and $\theta$ at $j_x=L_\mathrm{NCF}+L_\mathrm{DN}/2$.
    (e) $\bar{s}_{j_x,x}$,
    (f) $\bar{s}_{j_x,y}$, and
    (g) $\bar{s}_{j_x,z}$ are plotted as a function of $j_x$ for $(h/t,\theta/2\pi)=(0.5,0.1)$.
    $t\beta=10^3$, $L_y=40$, and $N_{\mathrm{sample}}=10^2$ samples averaged.
    }
    \label{fig:spin_SC_NC_DN}
\end{figure}

\begin{figure}[htbp]
    \centering
    \includegraphics[width=8.5cm]{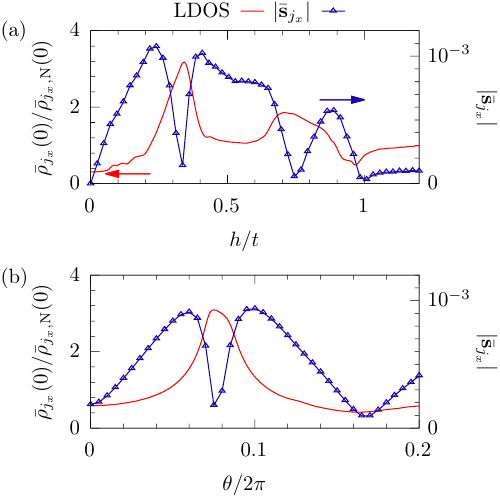}
    \caption{%
    The normalized zero energy LDOS (left $y$-axis) and $|\bar{\mathbf{s}}_{j_x}|$ (right $y$-axis) are plotted as a function of (a) $h/t$ for
    $\theta/2\pi=0.05$, and
    (b) $\theta/2\pi$ for
    $h/t=0.5$ at $j_x=L_\mathrm{NCF}+L_\mathrm{DN}/2$.
    $t\beta=10^3$, $L_y=100$, and $N_{\mathrm{sample}}=10^2$ samples averaged.
    }
    \label{fig:spin_cut_SC_NC_DN}
\end{figure}

We can see similar behaviors for the SC/NCF/DN junction (Figs.~\ref{fig:spin_SC_NC_DN} and \ref{fig:spin_cut_SC_NC_DN}).
In this case, $|\bar{\mathbf{s}}_{j_x}|$ also becomes small in the region where the normalized LDOS at zero energy exceeds unity
[Figs.~\ref{fig:spin_SC_NC_DN}(a) and \ref{fig:spin_cut_SC_NC_DN}].
Also, $\bar{\mathbf{s}}_{j_x}$ penetrates the DN due to the superconducting proximity effect [Figs.~\ref{fig:spin_SC_NC_DN}(e)--\ref{fig:spin_SC_NC_DN}(g)].


\bibliography{biblio}

\begin{thebibliography}{79}
\bibitem [{che()}]{chemical_potential}%
  \BibitemOpen
  \href@noop {} {}\bibinfo {note} {Here we consider a ballistic barrier because
  an insulating barrier greatly suppresses the amplitude of Cooper pair in the
  DN.}\BibitemShut {Stop}%
\end{thebibliography}
\end{document}